\documentclass[aps,prb,groupedaddress,twocolumn,10pt]{revtex4-1}

\usepackage{graphicx}

\begin{document}
\newcommand{\be}{\begin{equation}}
\newcommand{\ee}{\end{equation}}
\newcommand{\bea}{\begin{eqnarray}}
\newcommand{\eea}{\end{eqnarray}}
\newcommand{\nt}{\narrowtext}
\newcommand{\wt}{\widetext}

\title{Arrays of Josephson junctions between unconventional superconductors}

\author{D. V. Khveshchenko and R. Crooks}

\affiliation{Department of Physics and Astronomy, University of North Carolina, Chapel Hill, NC 27599}

\begin{abstract}
We study large arrays of mesoscopic junctions between gapless superconductors
where the tunneling processes of both, particle-hole and Cooper, pairs give rise
to a strongly retarded effective action which, contrary to the standard case, can not be readily
characterized in terms of a local Josephson energy. 
This complexity is expected to arise in, e.g., the grain boundary 
and $c$-axis junctions in layered high-$T_c$ superconductors.
A new representation for describing collective phenomena in this system is introduced,
and its phase diagram is discussed, alongside the electrical conductivity.
\end{abstract}

\maketitle

\nopagebreak

Quantum dynamics of ultrasmall normal and superconducting (Josephson) junctions (JJ)
has long been a field of active theoretical \cite{theory} and experimental \cite{exp}
research. Recently, the interest in this topic has been rekindled by a number of new ideas, such as the proposal 
of a novel 'floating' phase, in which context the effects of
(spatially) long-range correlations were investigated at a greater length \cite{tewari}.

Notably, though, most of the previous theoretical studies were limited to the
JJs between conventional, fully gapped, $s$-wave superconductors.
Although the case of the $d$-wave
superconducting cuprates, such as bi-epitaxial grain boundary (in-plain) JJs in $YBCO$ or
intrinsic $c$-axis (vertical) ones in $Bi2212$, have been rather extensively
studied as well, their previous analyses would routinely resort to a phenomenological
description similar to that of the gapped ($s$-wave) superconductors \cite{dwave}.
In contrast, the microscopic analysis of a single $d$-wave JJ carried out in Refs.\cite{joglekar,dvk},
showed that the processes of both, particle-hole and Cooper, pair tunneling can give
rise to the equally non-local (in the time domain) terms in the effective action,
thereby invalidating the very notion of a local Josephson energy.

In the present work, we study a large array of such JJs and
look into the effects of a strong time dependence of the effective action
on this system's phase diagram and corresponding behaviors.

The partition function of a generic single JJ introduced in Ref.\cite{theory}
can be generalized to the case of an array by including
both, self- and mutual-, capacitances

\begin{widetext}
\bea
S=\int^{1/T}_0d\tau{1\over 2}[\sum_iC_{ii}({\partial\phi_i(\tau)\over \partial\tau})^2
+\sum_{<ij>}C_{ij}({\partial\phi_{ij}(\tau)\over \partial\tau})^2]
\nonumber\\
-\sum_{<ij>}\int^{1/T}_0d\tau\int^{1/T}_0d\tau^\prime
[\alpha(\tau-\tau^\prime)\cos(\phi_{ij}(\tau)-\phi_{ij}(\tau^\prime))
+\beta(\tau-\tau^\prime)\cos(\phi_{ij}(\tau)+\phi_{ij}(\tau^\prime))],
\eea
\end{widetext}
where $\phi_{ij}(\tau)=\phi_i(\tau)-\phi_j(\tau)$ is the phase difference across the link $<ij>$.

The double time integrals in (1) are governed by the kernels
$\alpha(\tau)$ and $\beta(\tau)$ representing particle-hole and
Cooper pair tunneling processes, respectively. 
To the leading order in the tunneling matrix element
$T(k,k^\prime)$, they are given by the expressions
\be
\left(
\matrix{\alpha(\tau)\cr \beta(\tau)}
\right)=
-2\int {d^Dkd^Dk^\prime\over (2\pi)^{2D}}  |T(k,k^\prime)|^2
\left(
\matrix{{\cal G}_k(\tau){\cal G}_{k^\prime}(-\tau)\cr {\cal F}_k(\tau){\cal F}_{k^\prime}(-\tau)}
\right)\nonumber\\
\ee
where ${\cal G}$ and ${\cal F}$ are the normal and anomalous electron Green functions, respectively.

The $\alpha$-term describes (non-Gaussian) dissipation due to
the Andreev quasiparticle tunneling whose effects have been extensively
discussed in the previous works \cite{theory},
while the $\beta$-term represents the processes of (in general, non-synchronous)
pair tunneling. In the conventional ($s$-wave) superconductors,
it decays as $\beta(\tau)\propto e^{-\Lambda |\tau|}$,
thereby effectively reducing the last term in (1) to a single time integral
$E_J\int^{1/T}_0d\tau\cos 2{\phi_{ij}(\tau)}$ of what can then be
identified as the local Josephson energy $E_J=\int^{1/T}_0 d\tau\beta(\tau)$.

By contrast, in the case of a gapless superconductor one obtains strongly retarded
kernels \cite{joglekar,dvk}
\be
\alpha(\tau)/\alpha=\beta(\tau)/\beta=1/\tau^{2D-\eta}
\ee
where the prefactor in the $\beta$-kernel vanishes for any factorizable matrix element, 
$T(k,k^\prime)=f(k)f(k^\prime)$, of a symmetry other than the $s-wave$ and, therefore, 
it can only be due to a non-factorizable contribution   
$\propto f({\vec k}-{\vec k}^\prime)$ into $T(k,k^\prime)$.

In the two-dimensional case and under the condition of momentum conservation, 
$2D-\eta=2$ (see Refs.\cite{joglekar,dvk}), hence both the tunneling terms appear to be marginal, 
the corresponding coupling constants $\alpha$ and $\beta$ being dimensionless numbers of order unity.
A short-time divergence of Eqs.(3) can be naturally regularized by substituting $\tau\to
{\sqrt {\tau^2+\Lambda^{-2}}}$ where the cutoff scale $\Lambda$ is set by the maximal superconducting
gap in the bulk.

Conceivably, one can encounter even longer-ranged correlations
($2D-\eta<2$) in the presence of, e.g., resonant tunneling through zero energy states
supported by certain tunneling configurations, such as that of the $d_{0}/d_{\pi/4}$
in-plane grain boundary \cite{kawabata}.

Turning now to the effective action (1), we find that a strongly retarded nature of the 
tunneling terms renders a customary dual representation based on the Villain
transformation of the local Josephson term inapplicable, thereby making this model unsuitable
for the standard mapping onto an effective vortex plasma \cite{theory}.
Therefore, a well-known description of the different phases in terms of bound
vortex-antivortex complexes (dipoles, quadrupoles, etc.) can not be
readily generalized to the problem at hand, either, thus forcing one to take a different approach.

To that end, we introduce 
a new bosonic field $\psi_i(\tau)$, alongside an associated Lagrange multiplier field 
enforcing the local constraint $\psi_i(\tau)=e^{i\phi_i(\tau)}$. 
This approach should be contrasted with the previously developed
treatments of the conventional (local) Josephson term (see, e.g., Ref.\cite{kopec})
where a constrained bosonic variable would be used to represent the $pair$ field
$e^{2i\phi_i(\tau)}$. Indeed, an attempt to implement this technique in the present (non-local) case would
require one to work with a technically intractable bi-local composite operator $\psi_i(\tau)\psi_i(\tau^\prime)$.

By integrating out the phase variable $\phi_i$, keeping the leading
terms of the corresponding cluster expansion (cf. with Ref.\cite{kopec}), 
and then integrating out the Lagrange multiplier field, one arrives at the partition function
\begin{widetext}
\bea
Z=\int D\psi^\dagger_i(\tau) D\psi_i(\tau) D\lambda_i(\tau)
\exp(-\sum_{<ij>}\int^{1/T}_0d\tau_1\int^{1/T}_0d\tau_2
\psi^\dagger_i(\tau_1)[W_{ij}^{-1}(\tau_1-\tau_2)+\delta_{ij}\lambda_i(\tau_1)\delta(\tau_1-\tau_2)]
\psi_j(\tau_2)
\nonumber\\
+\alpha(\tau_1-\tau_2)
\psi^\dagger_i(\tau_1)\psi^\dagger_j(\tau_2)\psi_i(\tau_2)\psi_j(\tau_1)
+\beta(\tau_1-\tau_2)
\psi^\dagger_i(\tau_1)\psi^\dagger_i(\tau_2)\psi_j(\tau_2)\psi_j(\tau_1)+ h.c.)])
\eea
\end{widetext}
where $\lambda_i(\tau)$ is an additional Lagrange multiplier enforcing
the auxiliary constraint $\psi^\dagger_i(\tau)\psi_i(\tau)=1$
(the latter is not automatically satisfied, unless
the integration over $\phi_i(\tau)$ is performed exactly).

The correlation function appearing in Eq.(4)
\bea
W_{ij}(\tau)=\left<e^{i\phi_i(\tau)}e^{-i\phi_j(0)}\right>=\nonumber\\
\exp[-\int{d\omega d^Dk\over (2\pi)^{D+1}}{1-\cos(\omega\tau-{\vec k}{\vec R}_{ij})\over \omega^2C(k)}]=\delta_{ij}e^{-E_c|\tau|}\nonumber\\
\eea
is governed by the effective Coulomb energy $E_c=\int {d^Dk\over 2(2\pi)^{D+1}C_k}$
proportional to the integral of the inverse capacitance $C_k=\sum_{<ij>}C_{ij}e^{i{\vec k}{\vec R}_{ij}}$
which converges, provided that the capacitance matrix progressively decreases with the separation
between the sites.

The frequency integral in Eq.(5) diverges for any
${\vec R}_{ij}\neq 0$ which dictates that the correlation function $W_{ij}(\tau)$
remains strictly local in the real space.
Also, Eq.(5) is written in the limit of vanishing temperature, while at finite $T$
a proper account of large phase fluctuations with non-trivial winding numbers makes this
(as well as any bosonic) function periodic with a period $1/T$
by virtue of the substitution $\tau\to\tau- T\tau^2$ (see Ref.\cite{theory}).

At $\alpha=\beta=0$ one then obtains a bare (normal) Green function 
\be
G^{(0)}_{ij}(\omega)={2\delta_{ij}\over \omega^2/E_c+E_c},
\ee
while for finite $\alpha$ and $\beta$ the quantum charge fluctuations give rise to the corrections
which can be incorporated into the normal $G_{ij}=<\psi_i\psi^\dagger_j>$ and anomalous 
$F_{ij}=<\psi_i\psi_j>$ Green functions
obeying the usual Dyson's equations
\be
\left(
\matrix{G_{ij}\cr F_{ij}}
\right)
=
\left(
\matrix{G^{(0)}_{ij}\cr 0}
\right)
+G^{(0)}_{ik}
\sum_{kl}
\left(
\matrix{\Sigma_{kl} & \Delta_{kl}\cr
\Delta_{kl} & \Sigma_{kl}}
\right)
\left(
\matrix{G_{lj}\cr F_{lj}}
\right)
\ee
where both the normal $\Sigma_{ij}$ and anomalous $\Delta_{ij}$ self-energies 
can be computed as series expansions in powers of $\alpha$ and $\beta$.

The analysis of these expansions shows that they can be organized 
according to the powers of the inverse coordination number $z$ 
(e.g., $z=2D$ for a simple cubic lattice). In the leading 
approximation for $z\gg 1$, the self-energies are given by the equations

\bea
\Sigma_{ij}(\omega)=\int \frac{d\omega^\prime}{2\pi}[\delta_{ij}\sum_l
\alpha(\omega-\omega^\prime)G_{ll}(\omega^\prime)+ \nonumber\\
(\alpha(0)+\beta(0)+\beta(\omega-\omega^\prime))G_{ij}(\omega^\prime)]
\nonumber\\
\Delta_{ij}(\omega)=\int {d\omega^\prime\over 2\pi}
[\alpha(\omega-\omega^\prime)F_{ij}(\omega^\prime)+
\nonumber\\
\delta_{ij}\sum_l\beta(\omega-\omega^\prime)F_{ll}(\omega^\prime)]
\eea
When ascertaining a general layout of the phase diagram of the JJ array, 
different components of the self-energy can serve as emergent order parameters.
As such, one can distinguish between the local, $\Sigma_0=\Sigma_{ii}$, and non-local,
$\Sigma_1={1\over z}\sum_\mu\Sigma_{i,i+\mu}$ (here the sum is taken over the $z$ nearest neighbors), 
normal, as well the corresponding anomalous, $\Delta_0=\Delta_{ii}$ and $\Delta_1={1\over z}
\sum_\mu\Delta_{i,i+\mu}$, self-energies.

Specifically, $\Sigma_1$ signals the onset of a metallic
behavior (hopping between neighboring sites), $\Delta_0$ manifests an incipient local
pairing, $\Delta_1$ serves as the precursor of superconducting
coherence setting in across the entire JJ network, while a frequency-dependent part of the $\Sigma_0$
indicates the development of local time correlations.

With the on-site and nearest-neighbor terms taken into account, 
the spatial Fourier harmonics read 
\be
\left(
\matrix
{\Sigma(\omega,k) \cr
\Delta(\omega,k)}
\right)
=
\left(
\matrix
{\Sigma_{0}(\omega) \cr
\Delta_{0}(\omega)}
\right)
+
\left(
\matrix
{\Sigma_{1}(\omega) \cr
\Delta_{1}(\omega)}
\right)
\gamma(k)+...
\ee
where $\gamma(k)=\sum_{\mu} e^{ik\mu}$.

Eqs.(8) can be further improved by adding polarization corrections to the effective coupling terms
\be
\left(
\matrix
{\tilde\alpha \cr
\tilde\beta}
\right)
=
\left(
\matrix
{\alpha \cr
\beta}
\right)
+
\left(
\matrix
{\Pi_{E} & \Pi_{O} \cr
\Pi_{O} & \Pi_{E}}
\right)
\left(
\matrix
{\alpha & \beta \cr
\beta & \alpha}
\right)
\left(
\matrix
{{\tilde \alpha}\cr
{\tilde \beta}}
\right)
\ee
where the polarization functions $\Pi_{E,O}(\omega)=\int {d\omega^\prime\over 2\pi}\Gamma_{E,O}
G(\omega^\prime)G(\omega-\omega^\prime)$
include the vertex corrections $\Gamma_{E,O}$ arising from the 
even and odd numbers of non-crossing $\beta$-couplings
\be
\left(
\matrix{\Gamma_{E}\cr
{\Gamma_{O}}}
\right)
=
\left(
\matrix{1\cr \beta}
\right)
+
\left(
\matrix{\beta^{2} & 0\cr 0 & \beta^{2}}
\right)
\left(
\matrix{\Gamma_{E}\cr
{\Gamma_{O}}}
\right)
\ee
With the vertex and polarization corrections included and in the absence of any emergent order parameters,  
the self-consistent equation for $\Sigma_0(\omega)$ reads
\bea
\Sigma_0(\omega)=z\int {d\omega^\prime\over 2\pi}
{\bar \Gamma(\omega^\prime)}
{{\tilde \alpha}(\omega-\omega^\prime)\over G^{-1}_{0}(\omega^\prime)-\Sigma_{0}(\omega^\prime)}
\eea
The (static and spatially uniform) expectation value of the Lagrange multiplier
$\lambda=\left<\lambda_i(\tau)\right>$ can then be determined 
from the normalization condition $\int {d\omega d^Dk\over (2\pi)^{D+1}}G(\omega, k)=1$.

In order to ascertain the locations of the putative phase boundaries 
we include a constant term $\Sigma_0(0)+\lambda$ into the definition of the
renormalized Coulomb energy ${\tilde E}_c$
and expand Eqs.(8) to the first order in the emergent self-energies $\Sigma_1$, $\Delta_0$, $\Delta_1$, 
as well as the derivative of the (linear) frequency-dependent part of $\Sigma_0(\omega)$.
Threshold values of the couplings, beyond which
such self-energy components develop, are then given by the eigenvalue equations
\bea
\Sigma_1(\omega)=\int {d\omega^\prime\over 2\pi}\Gamma[{\tilde\alpha}(0)+{\tilde\beta}(0)+
{\tilde\beta}(\omega-\omega^\prime)]G^{2}_0(\omega^\prime)\Sigma_1(\omega^\prime)
\nonumber\\
\Delta_0(\omega)=z\int {d\omega^\prime\over 2\pi}
\Gamma{\tilde\beta}(\omega-\omega^\prime)
G^{2}_0(\omega^\prime)\Delta_0(\omega^\prime)
\nonumber\\
\Delta_1(\omega)=\int {d\omega^\prime\over 2\pi}\Gamma
{\tilde \alpha}(\omega-\omega^\prime)
G^{2}_0(\omega^\prime)\Delta_1(\omega^\prime)
\nonumber\\
{d\Sigma_0(\omega)\over d\omega}=
z\int {d\omega^\prime\over 2\pi}\Gamma{\tilde \alpha}(\omega^\prime)
G^{2}_0(\omega^\prime)
{d\Sigma_0(\omega^\prime)\over d\omega^\prime}
\nonumber\\
\eea
In the case of marginal ('Ohmic') dissipation corresponding to
$2D-\eta=2$ the Fourier transforms of the (regularized) coupling functions behave as
$\alpha(\omega)/\alpha=\beta(\omega)/\beta=\pi\Lambda e^{-|\omega|/\Lambda}$, thus resulting in
only a weak frequency dependence of the self-energy at $\omega\ll\Lambda$.

The first three of the eigenvalue equations (13) then reduce to the algebraic ones
\bea
1=(\Gamma_{E}^{2}+\Gamma_{O}^{2})(2\tilde\beta+\tilde\alpha)+
2\Gamma_{E}\Gamma_{O}(2\tilde\alpha+\tilde\beta)
\nonumber\\
1=z\left(\Gamma_{E}^{2}\tilde\beta+2\Gamma_{E}\Gamma_{O}\tilde\alpha+\Gamma_{O}^{2}\tilde\beta\right)
\nonumber\\
1=\Gamma_{E}^{2}\tilde\alpha+\Gamma_{O}^{2}\tilde\alpha+2\Gamma_{E}\Gamma_{O}\tilde\beta
\nonumber\\
\eea
from which one determines the locations of the putative critical lines in the $\alpha-\beta$ plane 
(see Fig1).

\begin{figure}
\includegraphics{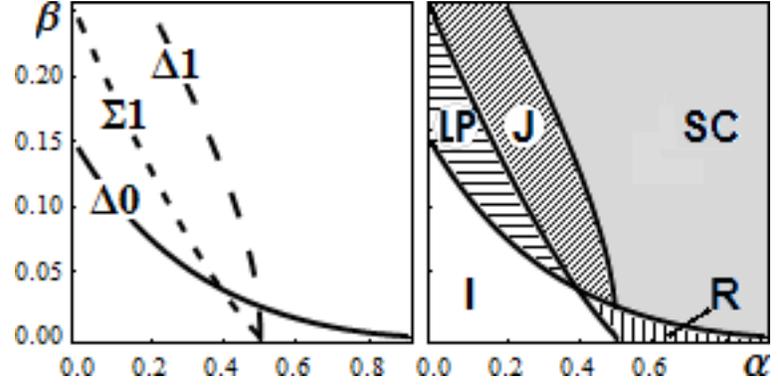}
\caption{Left panel: The onset of the inter-site self energy $\Sigma_{1}$ and 
both on-site and inter-site anomalous self energies $\Delta_0$ and $\Delta_1$. 
Right panel: phase diagram (see text)}
\end{figure}

Interestingly enough, Eqs.(14) suggest that for small $\alpha$ and large $z$ 
the onset of local ('on-site') pairing upon increasing $\beta$ may precede that of the metallic behavior,
while for small $\beta$ the inter-site ('bond') pairing emerges only at sufficiently large $\alpha$.

The above observations suggest a general layout of the phase diagram presented in 
Fig.1. The region of small $\alpha$ and $\beta$ with  
$\Sigma_1=\Delta_0=\Delta_1=0$ is interpreted as uniformly insulating (I),  while the emergent 
order parameter $\Delta_0\neq 0$ signals the onset of local superconducting pairing (LP)
at $\beta\sim 1/z$. At still higher values of $\beta\sim 1$ one expects to enter a Josephson-like 
phase (J) with $\Delta_0, \Sigma_1\neq 0$ but without global coherence. On the other hand, 
at $\alpha\sim 1$ the insulator gives way to the resistive phase (R) with  $\Sigma_1,\Delta_{1}\neq 0$
which supports both, Cooper pair and single quasiparticle, transport.
Lastly, the uniformly superconducting phase (SC) with $\Sigma_1,\Delta_{0,1}\neq 0$ would
eventually be attained at $\alpha, \beta\gtrsim 1$.
It should be noted, though, that our predictions are based on the approximate perturbative analysis and, therefore, 
not all the putative phase boundaries may actually be present in the real system. In particular, there may or may not 
be a physical distinction other than a crossover between the J and LP phases, or the latter regime 
might be absent altogether (as it is for $z=2$).

Such caveats notwithstanding, the overall behavior appears to be 
somewhat reminiscent of that in the standard ($s$-wave) case:
the system can be nudged closer to the superconducting state by increasing either, the Cooper pair
or particle-hole tunneling, the latter providing a mechanism for intrinsic dissipation which 
quenches phase fluctuations and promotes the classical Josephson effect.

Should, however, the tunneling $\beta$-term happen to decay even more slowly ($2D-\eta < 1$),
the analog of the effective Josephson energy would then diverge
at large $\tau$, thus making the infrared behavior essentially singular
and possibly allowing for some drastic changes in the phase structure.

Conducting properties of the JJ array allow one to discriminate between the different phases.
In particular, electrical conductivity can be computed as 
$\sigma_{\mu\nu}(\omega)={1\over i\omega}{\delta^2S[A]\over \delta A_\mu\delta A_\nu}$ with the use of the 
action of Eq.(1) in the presence of an external vector potential $A_\mu$, resulting in 
\begin{widetext}
\bea
\sigma_{\mu\nu}(\omega)=\int^{1/T}_0 d\tau
[\alpha(\tau){1-e^{i\omega\tau}\over \omega}\left<\cos(\nabla_\mu\phi(\tau)-\nabla_\nu\phi(0))\right>
+
\beta(\tau){1+e^{i\omega\tau}\over \omega}
\left<\cos(\nabla_\mu\phi(\tau)+\nabla_\nu\phi(0))\right>]+\dots
\eea
\end{widetext}
where the dots stand for 'paramagnetic' terms containing higher powers of $\alpha$ and $\beta$
which, therefore, are small compared to the above ('diamagnetic') contributions for $\alpha,\beta\lesssim 1$
(cf. with the discussion of a normal granular metal where $\beta=0$ in Ref.\cite{efetov}). 

The thus-obtained longitudinal conductivity reads  
\bea
\sigma_{\mu\mu}(\omega)\approx
\int^{1/T}_0 d\tau(\alpha(\omega){1-e^{i\omega\tau}\over \omega}[G^2_1(0)+G^2_0(\tau)+F_1^2(\tau)]+
\nonumber\\
\beta(\omega){1+e^{i\omega\tau}\over \omega}
[G^2_{1}(0)+G^2_1(\tau)+F_0^2(\tau)])
\nonumber\\
\eea
and, upon performing the frequency integrations, one obtains
\bea
\sigma_{\mu\mu}(\omega)\approx\alpha[{2E_c\over T}e^{-2E_c/T}(1+{\Delta^2_1\over E_c^2})
+{\Sigma^2_1\over E_c^2}]+
\beta\delta(\omega){\Sigma^2_1+\Delta^2_0\over E_c}
\nonumber\\
\eea
where, for the sake of simplicity, we chose $T\ll E_c=\Lambda$. 

The emergent metallicity order parameter $\Sigma_1$ promotes a metal-like  
(temperature-independent at $T\to 0$) conductivity, thereby distinguishing it from the
activation-type behavior characteristic of the insulating regime. Interestingly enough, 
it also contributes to the superfluid density, alongside the 
local pairing $\Delta_0$, while the non-local one 
($\Delta_1$) does not (to the lowest order in $\beta$).

It is conceivable, though, that there might be a (partial) cancellation
between the 'diamagnetic' and 'paramagnetic' terms at $\alpha,\beta\sim 1$,
as a result of which the conductivity could remain 
universal along the critical lines, akin to the situation in
the conventional, $s$-wave, JJ networks \cite{universal}
(it is worth reiterating that in the present case one can not readily invoke the charge-vortex
duality on which the universality argument is 
based \cite{theory} due to the inapplicability of the underlying Villain transformation).

To summarize, in the present work we studied arrays of unconventional JJs with long-range
(in the time domain) interactions stemming from the presence
of gapless quasiparticle excitations. On the technical side, the problem
presents a new challenge by not being amenable
to the customary approaches exploiting the intrinsic locality of 
the standard Josephson effective action.

By using an alternative representation, we find that the phase diagram of the 
system might feature the insulating, uniformly superconducting, Josephson (local pairing only), 
and metallic phases which can be identified by the corresponding emergent order parameters.
We also predict that this picture might be further altered in the presence 
of resonant tunneling between zero energy states 
where the temporal decay of correlations can be even longer-ranged.

We conclude by expressing a hope that this analysis will prompt a further
investigation into (and provide an alternative means for interpreting the experimental
data on) the assemblies of high-$T_c$ JJs beyond the scope of 
the customary phenomenological approach 
adapted from the earlier studies of the $s$-wave superconductors.


\end{document}